\documentclass[aps,prl,reprint,superscriptaddress,nofootinbib,floatfix]{revtex4-2}
\usepackage{graphicx,subfigure,bm,amssymb,amsmath,hyperref,dcolumn}
\usepackage{color,multirow,supertabular,float}
\usepackage{mathrsfs}
\usepackage{amsthm}
\usepackage{mathtools}
\usepackage{adjustbox}

\begin{document}

\title{Perturbative Renormalization and Universality Diagram for Long-Range Quantum Criticality}

\author{Zhiyi Li}
\affiliation{Department of Modern Physics, University of Science and Technology of China, Hefei, Anhui 230026, China}
\affiliation{Hefei National Laboratory, University of Science and Technology of China, Hefei 230088, China}

\author{Zhijie Fan}
\affiliation{Hefei National Laboratory, University of Science and Technology of China, Hefei 230088, China}
\affiliation{Shanghai Research Center for Quantum Science and CAS Center for Excellence in Quantum Information and Quantum Physics, University of Science and Technology of China, Shanghai 201315, China}
\affiliation{Hefei National Research Center for Physical Sciences at the Microscale and School of Physical Sciences, University of Science and Technology of China, Hefei 230026, China}

\author{Kun Chen}
    \email{chenkun@itp.ac.cn}
    \affiliation{CAS Key Laboratory of Theoretical Physics, Institute of Theoretical Physics, Chinese Academy of Sciences, Beijing 100190, China}

\author{Youjin Deng}
    \email{yjdeng@ustc.edu.cn}
\affiliation{Department of Modern Physics, University of Science and Technology of China, Hefei, Anhui 230026, China}
\affiliation{Hefei National Laboratory, University of Science and Technology of China, Hefei 230088, China}
\affiliation{Hefei National Research Center for Physical Sciences at the Microscale and School of Physical Sciences, University of Science and Technology of China, Hefei 230026, China}

\date{\today}

\begin{abstract}
Experimental progress in quantum simulators highlights the role of long-range (LR) interactions in reshaping quantum criticality and stabilizing exotic phases beyond the short-range (SR) paradigm.
We study ferromagnetic long-range quantum $O(n)$ models with interactions decaying as $1/r^{d+\sigma}$ and develop a perturbative renormalization-group expansion around the LR--SR boundary by setting $d=3-\epsilon$ and $\sigma=2-\delta$. 
In this parametrization, the full interacting LR window $2d/3<\sigma<2$ becomes $0<\delta<2\epsilon/3$, and is therefore perturbatively controlled. 
A two-loop calculation yields explicit expressions, in terms of $\epsilon$, $\delta$, and $n$, for the correlation-length exponent $\nu$ and for the frequency and momentum anomalous dimensions $\eta_\omega$ and $\eta_k$. 
The resulting exponents reduce to long-range Gaussian scaling at $\sigma=2d/3$ and to SR quantum Wilson-Fisher scaling in the $\sigma \to 2$ limit, thereby identifying $\sigma_*=2$ as the LR--SR boundary within the controlled $3-\epsilon$ expansion. 
Combining the RG results with scaling boundaries and classical LR analogies, we propose a $(d,\sigma)$ universality diagram for ferromagnetic long-range quantum $O(n)$ criticality and use it as an organizing framework for the phase diagram of long-range quantum spin chains.
\end{abstract}

\maketitle

\section{Introduction}

% Brief background of LR quantum criticality and related experimental progress.
Long-range (LR) interactions can modify quantum many-body physics by directly coupling distant degrees of freedom.
Algebraically decaying interactions can produce phases and critical regimes absent from short-range (SR) models, including nonlocal correlation propagation, conformal-symmetry breaking, and long-range-induced topological phases and phase transitions~\cite{Defenu2023,Hauke2013,Schachenmayer2013,Eisert2013,Richerme2014,Jurcevic2014,Vodola2014,Lepori2016,Gong2023,LeporiDellAnna2017}.
Experimental platforms, including cold atoms with dipolar or Rydberg-dressed interactions, trapped-ion chains with power-law spin-spin couplings, and Rydberg atom arrays, provide direct means to implement and tune such interactions~\cite{Lahaye2009,Britton2012,Barredo2016,Bernien2017,Browaeys2020}.
These developments motivate a systematic theory of quantum criticality in models with interactions decaying as $1/r^{d+\sigma}$.

% Do we really need the Hamiltonian here? This equation is never referred to. I will keep it here for now. 
% 干脆不要了？ 还能缩短长度。
The long-range quantum $O(n)$ spin model is a standard setting for this question, with microscopic Hamiltonian
\begin{equation}
H = -\frac{1}{2}\sum_{i\ne j}J(r_{ij})
\mathbf n_i\cdot\mathbf n_j 
+ \Gamma \hat H_n
\label{eq:rotor}
\end{equation}
where $\mathbf n_i$ are $n$-component unit vectors on site $i$, and $J(r_{ij}) = 1/|r_i-r_j|^{d+\sigma}$ denotes the algebraically decaying LR interaction. Here, $\Gamma$ controls quantum fluctuations, and $\hat H_n$ denotes the transverse-field term for $n=1$ (Ising) and the angular-momentum term for $n>1$ (the XY rotor model when $n=2$, and the Heisenberg model when $n=3$).

% Start just here is also acceptable
At long distances, the model can be described by a Euclidean continuum field theory in spatial and imaginary-time coordinates $(\mathbf{x}, \tau)$,
\begin{multline}
S=\int d\tau d^d \mathbf{x}\,
\frac{1}{2}\boldsymbol{\phi}_{\mathbf{x},\tau}
\bigl[-\partial_\tau^2+(-\Delta)^{\sigma/2}+K(-\Delta)+r\bigr]\boldsymbol{\phi}_{\mathbf{x},\tau}
\\
+\frac{u}{4!}\int d\tau d^d\mathbf{x}\,(\boldsymbol{\phi}_{\mathbf{x},\tau} 
^2)^2,
\label{eq:action}
\end{multline}
% ZJ: need to write the correct $n$-compnent form.
Here, $\boldsymbol{\phi}(\mathbf{x},\tau)$ is the $n$-component vector field, and the fractional Laplacian $(-\Delta)^{\sigma/2}$ encodes the nonanalytic LR kernel $|\mathbf{k}|^\sigma$. The analytic Laplacian $-\Delta$ term represents the short-range (SR) $k^2$ operator generated under coarse-graining, with $\mathbf{k}$ denoting the corresponding momentum~\cite{nijboer1957calculation}. 
At the quantum critical point $r = 0$, the frequency operator $\omega^2$ and the dominant momentum operator $|\mathbf{k}|^\sigma$ renormalize differently, so the quantum critical point is characterized by two independent anomalous dimensions, $\eta_\omega$ and $\eta_k$.
Consequently, the dynamical exponent $z$ is related to these two anomalous dimensions through $z = (2-\eta_k)/(2-\eta_\omega)$.

Despite the simple form of Eq.~\eqref{eq:action}, a controlled field-theoretical description of this model is still lacking. 
Existing studies have probed long-range quantum criticality through various field-theoretic and numerical approaches. 
Early one-loop renormalization-group (RG) analyses mapped the overall phase structure and identified the LR Gaussian regime for $\sigma < 2d/3$, the interacting LR regime for $\sigma \in (2d/3,2)$, and the SR quantum critical regime for $\sigma > 2$~\cite{DuttaBhattacharjee2001}. 
Two-loop RG calculations based on an $\epsilon'=3\sigma/2-d$ expansion near the upper critical surface $d_u = 3\sigma/2$ obtained a correction to $\eta_\omega$ while treating $\eta_k$ at its mean-field value. These calculations therefore derived a correction to the dynamical exponent $z$, but they do not explain how $z$ reduces to $1$ when approaching the SR regime~\cite{Maghrebi2016, Benedetti2024Dynamic}.
By analogy with classical long-range systems, the LR--SR boundary of quantum LR systems has also been proposed to follow a Sak-type shift at $\sigma^*=2-\eta_{\rm SR}$, where $\eta_{\rm SR}$ is the anomalous dimension in the SR limit~\cite{Sak1973,Maghrebi2016,Defenu2017,Defenu2023}.
% ? if there are two anomalous dimensions $\eta_k$ and $\eta_\omega$, the Sak shift might results in a complicated boundary or requires $z = 1$ in SR limit.
Existing numerical studies using tensor-network methods, quantum Monte Carlo (QMC), and high-order linked-cluster methods have mapped out the principal mean-field, continuously varying LR, and SR regimes of the one-dimensional (1D) LR transverse-field Ising model. A controlled description of the LR--SR boundary near $\sigma\simeq2$ and of the critical exponents remains incomplete~\cite{Koffel2012,Jaschke2017,Zhu2018,Koziol2021,FeySchmidt2016,Fey2019}. For the 1D Bose-Hubbard model with long-range hopping, recent QMC simulations have also established continuous superfluid--Mott criticality that is not of the Berezinskii--Kosterlitz--Thouless (BKT) type for $0<\sigma\le2$~\cite{Gupta2026BoseHubbard}. These results call for a controlled field-theoretic description that can access the entire interacting LR window.

Inspired by recent work on the $4-\epsilon$ expansion for classical LR criticality~\cite{LongRangeBoundary2026}, we perform an RG analysis of the quantum LR model using a controlled expansion parametrized by $d = 3-\epsilon$ and $\sigma = 2-\delta$. This parametrization maps the full non-mean-field LR window to $0<\delta<2\epsilon/3$. Hence the whole non-mean-field LR region falls within the perturbative regime for small $\epsilon$: throughout this window $\delta=O(\epsilon)$, so the expansion remains controlled. At two-loop order, we identify a nontrivial LR Wilson-Fisher (WF) point, at which the correlation-length exponent and the frequency and momentum anomalous dimensions are given by
\begin{subequations}
    \label{criticalexponent}
\begin{align}
\frac{1}{\nu} &=\sigma-\frac{n+2}{n+8}\left(\epsilon-\frac{3\delta}{2}\right) +O(\epsilon^2) \label{eq:nu} \\
\eta_\omega&=\frac{n+2}{2(n+8)^2}\left(\epsilon-\frac{3}{2}\delta\right)^2+O(\epsilon^3) \label{eq:etaomega}
\\
\eta_k&=\delta+\frac{n+2}{2(n+8)^2}\frac{\left(\epsilon-\frac{3}{2}\delta\right)^3}{\epsilon-\delta}+O(\epsilon^3).
\label{eq:etak}
\end{align}
\end{subequations}
These results imply the dynamical exponent
\begin{equation}
z=1-\frac{\delta}{2}
+
\frac{(n+2)\delta}{8(n+8)^2(\epsilon-\delta)}
\left(\epsilon-\frac{3}{2}\delta\right)^2
+O(\epsilon^3).
\label{eq:zmain}
\end{equation}
These expressions reduce to the short-range quantum WF values at $\sigma=2$ and to the long-range Gaussian values at $\sigma=2d/3$. They also recover the exact results for the spherical model in the $n \to \infty$ limit. This consistency suggests that $\sigma^*=2$ is the LR--SR boundary for the LR quantum $O(n)$ model within the controlled $3-\epsilon$ expansion.

\begin{figure}[t]
\centering
\includegraphics[width=\columnwidth]{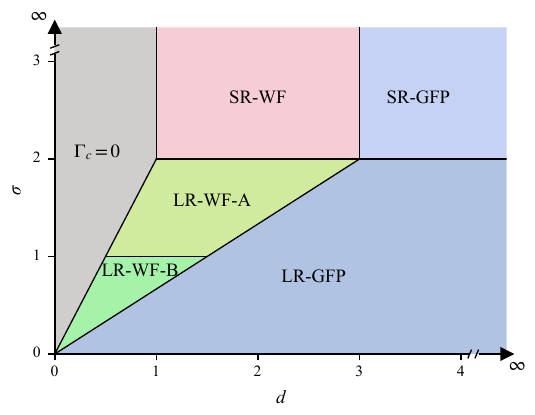}
\caption{Proposed universality diagram for quantum long-range $O(n)$ spin models in the $(d,\sigma)$ plane. The colored regions denote distinct zero-temperature quantum critical regimes: the region with no finite-field quantum criticality ($\Gamma_c=0$), the LR Gaussian regime (LR-GFP), the interacting LR Wilson-Fisher sectors (LR-WF-A and LR-WF-B), the SR Wilson-Fisher regime (SR-WF), and the SR Gaussian regime (SR-GFP). Solid lines indicate the main boundaries of the diagram. The lower and upper critical boundaries are scaling boundaries, with the SR lower and upper critical dimensions included on the short-range side; the $\sigma=2$ line is the RG-supported LR--SR boundary; and the line at $\sigma=1$ marks the conjectured subdivision of the interacting LR-WF regime.}
\label{fig:universality}
\end{figure}

Using the RG results together with scaling boundaries and classical LR analogies, we propose a $(d,\sigma)$ universality diagram for quantum long-range $O(n)$ spin models, as shown in Fig.~\ref{fig:universality}. The lower and upper critical boundaries follow from scaling, the LR--SR boundary at $\sigma=2$ is supported by the RG analysis, and the internal line at $\sigma=1$ is motivated by long-range simple-random-walk and LR-percolation analogies. We then use the $d=1$ cross section to organize the phase diagrams of the representative long-range quantum spin chains.

\section{Perturbative Renormalization Group Analysis}
The continuum action in Eq.~\eqref{eq:action} provides the starting point for the perturbative renormalization-group theory. We begin with the tree-level analysis by assigning canonical dimensions $[k]=1$ and $[\omega]=z$.
The full quadratic inverse bare propagator is $G_0^{-1}(\omega,k)=\omega^2+|\mathbf{k}|^\sigma+Kk^2+r$.
The analytic gradient $Kk^2$ has tree-level eigenvalue $y_{k^2}=\sigma-2$, relative to the nonanalytic kernel $|\mathbf{k}|^\sigma$.
For $\sigma>2$, the $Kk^2$ term dominates over $|\mathbf{k}|^\sigma$ and drives the theory to the SR universality class. For $\sigma<2$, the $Kk^2$ term is irrelevant, and we can set $K=0$ and retain only the LR kinetic term.
Thus, at tree level, the LR--SR boundary is $\sigma=2$.
In the LR regime, matching $\omega^2$ and $|\mathbf{k}|^\sigma$ fixes $z_0=\sigma/2$, and the mass eigenvalue is $y_r^{(0)}=\sigma$.

\begin{figure}[t]
\centering
\includegraphics[width=\linewidth]{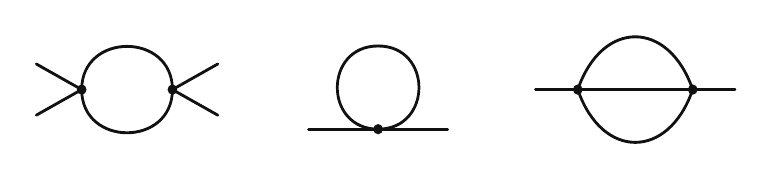}
\caption{The one-loop bubble (left) determines the beta function and the WF fixed point.
The one-loop tadpole (middle) renormalizes the mass and gives the leading correction to $y_r=1/\nu$.
The two-loop sunset (right) produces the leading kinetic counterterms; the frequency and momentum projections of this graph carry distinct pole structures.
All notation is explained in the text and the Supplemental Material.}
\label{fig:feynman}
\end{figure}

The lower critical dimension can be obtained from the critical bare propagator at $r=0$. The fluctuation integral behaves as 
$\int d\omega\, d^d k\,  G_0(\omega,k) \sim \int_0^\Lambda dk\, k^{d-1-\sigma/2}$, where $\Lambda$ is the ultraviolet cutoff.
The integral is infrared divergent for $d\le\sigma/2$, so long-wavelength fluctuations destroy the finite-field ordered phase and yield $d_\ell=\sigma/2$.
The upper critical boundary is instead determined by the relevance of the interaction.
Requiring the action to be dimensionless gives the canonical field scaling, $[\phi(\mathbf{x},\tau)]=(d-\sigma/2)/2$.
The quartic coupling $u$ in Eq.~\eqref{eq:action} then has canonical dimension $y_u^{(0)}=3\sigma/2-d$, which gives the LR upper critical dimension $d_u=3\sigma/2$.
For $d>d_u$, the quartic interaction is irrelevant, and the transition is governed by the noninteracting LR Gaussian theory, giving the LR mean-field regime.
For $d<d_u$, interactions become relevant and an interacting fixed point is required.

For the perturbative RG analysis, we organize the theory as a $3-\epsilon$ expansion with $d=3-\epsilon$. In the full non-mean-field LR window, $2d/3<\sigma<2$, writing $\delta=2-\sigma$ gives $0<\delta<2\epsilon/3$. Thus $\delta=O(\epsilon)$ throughout the window, and the whole LR-WF regime is perturbatively controlled within the same $\epsilon$ expansion.

We first consider the one-loop calculation.
The dimensionless renormalized coupling $g$ is defined by absorbing the one-loop bubble residue into $u$; with this normalization, the four-point bubble in Fig.~\ref{fig:feynman} (left) gives the beta function
\begin{equation}
\beta_g\equiv \mu \frac{d g}{d\mu}
=-\left(\epsilon-\frac{3\delta}{2}\right) g+\frac{n+8}{6}g^2+O(g^3)\label{eq:betag}.
\end{equation}
Here $\mu$ is the RG scale.
Solving $\beta_g(g_*)=0$, we find a fixed point at
\begin{equation}
g_*=\frac{6}{n+8}\left(\epsilon-\frac{3\delta}{2}\right)+O(\epsilon^2).
\end{equation}
Thus, $y_u = -\beta_g'(g_*)=-(\epsilon-3\delta/2)+O(\epsilon^2)<0$, showing the infrared (IR) stability of the fixed point.

The one-loop tadpole self-energy in Fig.~\ref{fig:feynman} (middle) renormalizes the tuning parameter $r$ and gives $\beta_r\equiv \mu\,dr/d\mu=-(\sigma-(n+2)g/6)r+O(g^2r)$.
At the fixed point, $r^*=0$, and the corresponding relevant eigenvalue is $y_r \equiv \nu^{-1} =-\beta'_r(g_*)=\sigma-(n+2)g_*/6+O(\epsilon^2)$, as shown in Eq.~\eqref{eq:nu}.

The renormalization of the kinetic terms first appears at two loops, since the one-loop tadpole is independent of the external frequency and momentum.
The relevant contribution is the sunset self-energy in Fig.~\ref{fig:feynman} (right).
For this nonlocal theory with a $|\mathbf{k}|^\sigma$ kinetic term in the bare action, we take the kinetic counterterms of the form
\begin{equation}
\delta\Gamma^{(2)}_{\rm kin}(\omega,k)
=\delta Z_\omega\,\omega^2+\delta Z_k\,|\mathbf{k}|^\sigma .
\label{eq:kineticcounterterm}
\end{equation}
The RG conditions are imposed by requiring the renormalized inverse propagator to have finite coefficients of both $\omega^2$ and $|\mathbf{k}|^\sigma$.
% \begin{equation}
%     \delta Z_\omega=\left .\frac{\partial \Sigma}{\partial \omega^2}\right|_{\omega=0,k=0}, \quad
%     \delta Z_k=\left .\frac{\partial \Sigma}{\partial |\mathbf{k}|^\sigma}(\omega,k)\right|_{\omega=0,k=0}.
% \end{equation}
Specifically, the projection of the sunset graph onto the external frequency term gives the frequency counterterm
\begin{equation}
    \delta Z_\omega=\left .\frac{\partial \Sigma (\omega,k)}{\partial \omega^2}\right|_{\omega=0,k=0}. \label{eq:deltaZomega}
\end{equation}
The logarithmic divergence of this projection is governed by $\Gamma[(2\epsilon-3\delta)/2]$.
Meanwhile, projecting the same graph onto the nonanalytic external momentum structure $|\mathbf{k}|^\sigma$ gives the spatial counterterm
\begin{equation}
    \delta Z_k=\left .\frac{\partial \Sigma(\omega,k)}{\partial |\mathbf{k}|^\sigma}\right|_{\omega=0,k=0}, \label{eq:deltaZk}
\end{equation}
whose singularity is governed by $\Gamma(\epsilon-\delta)$. See the Supplemental Material for detailed calculations. Thus, the frequency and momentum channels contain different pole structures, proportional to $1/(2\epsilon-3\delta)$ and $1/(\epsilon-\delta)$, and require independent kinetic renormalizations.

% validity of RG
By comparison, the conventional expansion in $\epsilon'=3\sigma/2-d$ is constructed near the LR upper critical dimension and controls the departure from mean-field behavior.
It does not perturbatively address the approach to the LR--SR boundary at $\sigma=2$.
In this setting, the two-loop momentum projection does not generate a $1/\epsilon'$ divergence, so previous field-theoretic analyses found no need to introduce a spatial fractional-kernel counterterm $\delta Z_k$ at this order, leaving $\eta_k$ unrenormalized beyond its mean-field value~\cite{Benedetti2024Dynamic}.
In the present $3-\epsilon$ expansion, the full non-mean-field LR window maps to $0<\delta<2\epsilon/3$ with $\delta=2-\sigma=O(\epsilon)$, so the approach to $\sigma=2$ remains within the controlled perturbative regime.
The momentum projection then retains the pole at $\epsilon-\delta$, requiring $\delta Z_k$ and producing the loop correction to the spatial anomalous dimension in Eq.~\eqref{eq:etak}.

The anomalous dimensions are obtained from the Callan--Symanzik equation using the kinetic counterterms determined in minimal subtraction (MS): the loop contribution in each channel is $-\beta_g\partial_g\delta Z_\alpha$ evaluated at $g_*$, with $\alpha=\omega,k$.
This gives $\eta_\omega$ in Eq.~\eqref{eq:etaomega}; for the spatial exponent, our convention $\Gamma^{(2)}(0,k)\sim |k|^{2-\eta_k}$ adds the canonical LR contribution $\delta$ from $|\mathbf{k}|^\sigma=|k|^{2-\delta}$, yielding Eq.~\eqref{eq:etak}.
The dynamical exponent then follows from the homogeneous scaling relation $z=(2-\eta_k)/(2-\eta_\omega)$, giving Eq.~\eqref{eq:zmain}.
The limiting cases are consistent with this picture.
At $\epsilon-3\delta/2=0$, or $\delta=2\epsilon/3$, the fixed point coupling $g_*$ vanishes and all exponents reduce to LR Gaussian mean-field behavior.
At $\delta=0$, the LR kernel becomes the analytic $k^2$ kernel, Eqs.~\eqref{eq:etaomega} and~\eqref{eq:etak} give $\eta_\omega=\eta_k$, and Eq.~\eqref{eq:zmain} gives $z=1$, recovering the SR quantum Wilson-Fisher universality class.
Therefore, the RG analysis connects the LR mean-field boundary to the SR boundary in the controlled $3-\epsilon$ perturbative expansion and supports $\sigma_*=2$ as the LR--SR boundary.

\section{Universality Diagram}
% 按“scaling-derived boundaries → RG-derived regions → conjectural A/B subdivision → table”组织。

The universality of quantum long-range $O(n)$ spin models at a given spatial dimension $d$ and decay exponent $\sigma$ depends on the relevance of the quartic interaction, the competition between the nonanalytic LR kinetic term and the analytic SR kinetic term, and the lower critical boundary for finite-field quantum criticality. Taking these factors into account, and guided by Ref.~\cite{XiaoFanDeng2026}, we propose a universality diagram for the long-range quantum $O(n)$ model in the $(d,\sigma)$ plane, shown in Fig.~\ref{fig:universality}.

The universality diagram is organized by three boundaries: the lower and upper critical dimensions and the LR--SR boundary. 
The lower and upper boundaries are obtained by matching the LR scaling results to the corresponding SR quantum critical dimensions.
On the LR side, scaling gives $d_\ell=\sigma/2$ and $d_u=3\sigma/2$. On the SR side, the corresponding quantum $O(n)$ theory has lower and upper critical spatial dimensions $d_\ell^{\rm SR}=1$ and $d_u^{\rm SR}=3$. Combining the two sides gives $d_\ell=\min(\sigma/2,1)$, below which no finite-field quantum phase transition is expected, and $d_u=\min(3\sigma/2,3)$, above which the quartic interaction becomes irrelevant and the transition is governed by the corresponding Gaussian fixed point. The LR--SR boundary at $\sigma=2$ is fixed by the competition between the LR and SR kinetic terms and is supported by the $3-\epsilon$ RG analysis. This boundary is also consistent with analogous results for classical LR systems~\cite{XiaoXY2025, YaoXY2025, xiaoSpontaneousSymmetryBreaking2026, LiuXiaoFanDeng2025PercolationPrep}.

In addition to the scaling- and RG-derived boundaries, we mark $\sigma=1$ as a conjectured internal boundary within the interacting LR-WF regime. 
The motivation comes from long-range simple-random-walk (LR-SRW) arguments and from the analogy with LR percolation~\cite{XiaoFanDeng2026}. 
In LR-SRW, $\sigma=1$ marks the onset of hyperballistic transport: for $\sigma<1$, the first moment of the step distribution diverges, so rare long jumps strongly reshape the effective spatial metric~\cite{BergersenRacz1991, JanssenOerdingVanWijlandHilhorst1999}. 
Recent mathematical results for 2D LR percolation further show that, within $2/3 < \sigma < 1$, the anomalous dimension takes the LR-GFP value, $\eta=2-\sigma$, while other critical exponents remain nontrivial~\cite{Hutchcroft2025Dimension, Hutchcroft2025CriticalI, Hutchcroft2025CriticalII, Hutchcroft2025CriticalIII}; this picture is consistent with recent numerical results~\cite{LiuXiaoFanDeng2025PercolationUnpublished}. 
This motivates the LR-WF-B regime ($\sigma < 1$ and $d \in (\sigma/2, 2\sigma /3)$) in the quantum problem: $\nu$ and $\eta_\omega$ are governed by the interacting LR-WF fixed point, whereas the momentum anomalous dimension retains the LR-GFP value, $\eta_k=2-\sigma$, reflecting the strongly nonlocal spatial kernel.

The resulting diagram contains six regimes: the region with no finite-field quantum criticality; the LR Gaussian regime (LR-GFP); two interacting LR critical regimes, LR-WF-A and LR-WF-B; and the SR Wilson-Fisher (SR-WF) and SR Gaussian regimes (SR-GFP). The corresponding critical exponents are summarized in Table~\ref{tab:region_exponents}. The LR-WF-A entries use the two-loop RG expressions in Eq.~\eqref{criticalexponent}, whereas the LR-WF-B row should be read as the conjectured scenario described above.

\begin{table}[t]
\caption{Critical-exponent behavior for the regions in Fig.~\ref{fig:universality}. Here, the LR-WF critical exponents are denoted by $y_r^{\rm LR},\eta_\omega^{\rm LR},\eta_k^{\rm LR}$, with the two-loop perturbative RG results given in Eq.~\eqref{criticalexponent}. The SR-WF critical exponents are denoted by $y_r^{\rm SR},\eta^{\rm SR}$, with the two-loop perturbative RG results given as $y_r^{\rm SR}=2-\frac{n+2}{n+8}\epsilon+O(\epsilon^2)$ and $\eta^{\rm SR}=\frac{n+2}{2(n+8)^2}\epsilon^2+O(\epsilon^3)$. The LR-WF-B assignment for $\eta_k$ is conjectural and is motivated by the random-walk/percolation analogy discussed in the text.}
\label{tab:region_exponents}

\begin{tabular}{l|c@{\hspace{4.2em}}c@{\hspace{3.2em}}c}
\hline\hline

Region & $\nu^{-1}$ & $\eta_\omega$ & $\eta_k$ \\
\hline
LR-GFP & $\sigma$ & $0$ & $2-\sigma$ \\
LR-WF-A & $y_r^{\rm LR} $ & $\eta_\omega^{\rm LR}$ & $\eta_k^{\rm LR}$ \\
LR-WF-B & $y_r^{\rm LR} $ & $\eta_\omega^{\rm LR}$ & $2-\sigma$ \\
SR-WF & $y_r^{\rm SR}$ & $\eta^{\rm SR}$ & $\eta^{\rm SR}$ \\
SR-GFP & $2$ & $0$ & $0$ \\
$\Gamma_c=0$ &  \multicolumn{3}{c}{no finite-field quantum criticality} \\
\hline\hline

\end{tabular}
\end{table}

\section{Phase Diagram of the $d=1$ Case}
% 按“zero-$T$ quantum transitions → finite-$T$ transitions → model dependence”组织。

\begin{figure*}[t]
    \centering
    \includegraphics[trim={0cm 0.2cm 0cm 0cm}, clip, width=\linewidth]{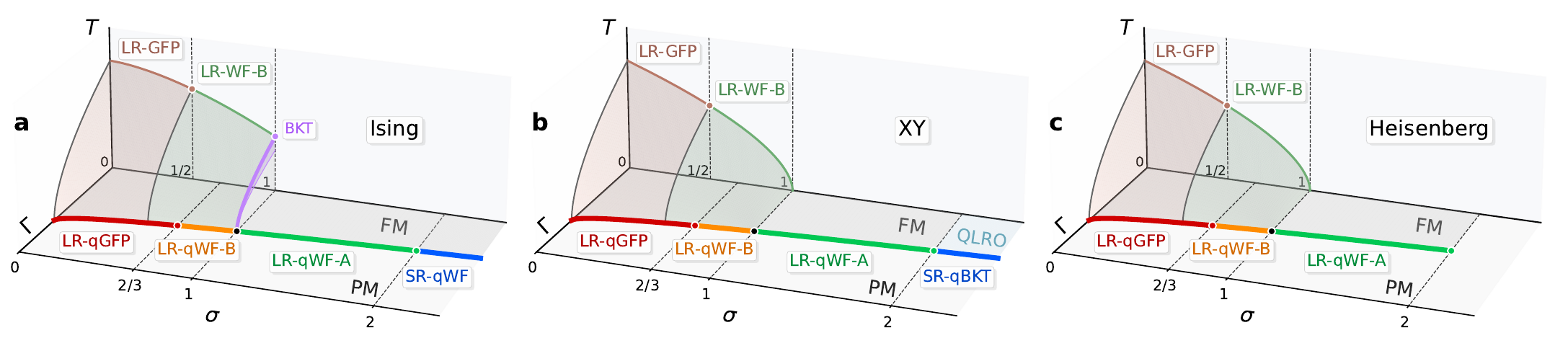}
    \caption{Schematic temperature--quantum-coupling phase diagrams of one-dimensional long-range quantum spin chains:
    (a) transverse-field Ising, (b) quantum XY/rotor, and (c) quantum Heisenberg.
    The axes are temperature $T$, quantum coupling $\Gamma$, and decay exponent $\sigma$.
    Colored lines in the $T=0$ plane denote quantum critical regimes: LR-qGFP for $0<\sigma<2/3$, LR-qWF-B for $2/3<\sigma<1$, LR-qWF-A for $1<\sigma<2$, and the appropriate short-range quantum regime for $\sigma>2$.
    The prefix ``q'' distinguishes zero-temperature quantum fixed points from finite-temperature classical long-range regimes.
    For $\sigma<1$, the finite-temperature critical surfaces are inherited from the corresponding classical one-dimensional long-range models, with LR-GFP and LR-WF-B sectors separated at $\sigma=1/2$.
    At $\sigma=1$, the Ising chain has a BKT-like finite-temperature boundary.
    For $\sigma>2$, the Ising chain is governed by SR-qWF criticality, the XY/rotor chain by SR-qBKT criticality, and the Heisenberg chain has no finite-$\Gamma$ transition.}
    \label{fig:d1phase}
\end{figure*}

One-dimensional long-range quantum spin chains are relevant to several current experimental platforms~\cite{Britton2012,richermeNonlocalPropagationCorrelations2014, Jurcevic2014, lagoinBoseHubbardSimulator2026,morinSupersolidCrystalsDipolar2025, sundaresanInteractingQubitphotonBound2019, zhangSuperconductingQuantumSimulator2023}. 
We use the $d=1$ cross section of the $(d,\sigma)$ universality diagram to construct the $(T,\Gamma,\sigma)$ phase diagrams in Fig.~\ref{fig:d1phase} for three representative models: the transverse-field Ising chain, the quantum XY rotor chain, and the quantum Heisenberg chain. 
The colored critical lines in the $T=0$ plane denote quantum critical regimes inherited from Fig.~\ref{fig:universality}, while the finite-temperature surfaces follow the corresponding one-dimensional classical long-range critical behavior~\cite{DuttaBhattacharjee2001, Defenu2017, XiaoFanDeng2026}.

At $T=0$, the upper critical boundary occurs at $\sigma=2/3$. 
For $0<\sigma<2/3$, the transition is governed by the LR quantum Gaussian fixed point. 
For $2/3<\sigma<2$, it is governed by an interacting LR quantum fixed point. 
Within this interacting LR interval, we retain the conjectured subdivision introduced above: $2/3<\sigma<1$ is labeled LR-qWF-B, while $1<\sigma<2$ is labeled LR-qWF-A. 
For $\sigma>2$, the analytic $k^2$ kinetic term takes over and the system enters the short-range quantum regime. 
The boundary point $\sigma=2$ is assigned to the LR side of this classification, consistent with numerical studies of two-dimensional LR interacting classical systems~\cite{XiaoXY2025, YaoXY2025, LiuXiaoFanDeng2025PercolationPrep}.

On the short-range side, the long-range tail is irrelevant and the quantum behavior becomes model dependent, as in the corresponding two-dimensional SR classical models. 
The Ising chain has the usual ferromagnet--paramagnet quantum phase transition governed by the SR quantum Wilson-Fisher fixed point. 
The quantum XY rotor chain instead undergoes a BKT transition between a paramagnetic phase and a low-$\Gamma$ quasi-long-range ordered phase. 
The Heisenberg chain has no finite-$\Gamma$ quantum phase transition in the short-range limit.

The finite-temperature part of the diagram is inherited from the corresponding one-dimensional classical long-range models. 
In this classical problem, $\sigma=1/2$ separates the LR Gaussian and interacting LR regimes, while $\sigma=1$ is the marginal lower-critical boundary~\cite{Dyson1969,Thouless1969,Kosterlitz1976,LuijtenMessingfeld2001,XiaoFanDeng2026}. 
For the Ising chain, the inverse-square case at $\sigma=1$ gives a finite-temperature BKT-like transition associated with logarithmically interacting domain-wall excitations, below which lies a finite regime with divergent susceptibility~\cite{Thouless1969, Kosterlitz1976, FukuiTodo2009, Humeniuk2020, knapProbingRealspaceTimeresolved2013}. 
For $\sigma>1$ in the Ising chain, and for $\sigma\ge1$ in the XY and Heisenberg chains, no finite-temperature transition is expected.

\section{Discussion and Outlook}

We have developed a controlled $3-\epsilon$ perturbative RG framework for quantum long-range $O(n)$ criticality by setting $d=3-\epsilon$ and $\sigma=2-\delta$.
In this formulation, the full interacting LR window has $\delta=O(\epsilon)$, so the same expansion describes both the approach to the LR Gaussian boundary and the approach to the SR regime.
The two-loop calculation yields independent anomalous dimensions in frequency and momentum, $\eta_\omega$ and $\eta_k$, and connects to the expected limiting behaviors at $\sigma=2d/3$ and $\sigma=2$.
These results provide analytic support for the LR--SR boundary $\sigma_*=2$ within the controlled expansion.

Together with scaling arguments and classical LR analogies, the RG results motivate a $(d,\sigma)$ universality diagram for long-range quantum criticality. In this diagram, the lower and upper critical boundaries come from scaling, the LR--SR boundary is supported by the RG calculation, and the $\sigma=1$ subdivision is motivated by LR-SRW and LR-percolation analogies. The $d=1$ cross section then organizes the zero-temperature quantum critical regimes and their connection to finite-temperature behavior in long-range spin chains.

Further work should include higher-loop field-theory calculations, extensions to additional observables, and high-precision numerical studies of long-range quantum spin models.
Such studies can test the predicted exponents, the RG-supported boundary $\sigma_*=2$, and the conjectured $\sigma=1$ subdivision of the interacting LR-WF regime.

\begin{acknowledgments}
The authors thank Tianning Xiao for valuable discussions.
K.C. is supported by the National Key Research and Development Program of China, Grant No. 2024YFA1408604, the National Natural Science Foundation of China under Grant Nos. 12474245 and 12447103, and the GHfund A (202407010637).
Z.L., Z.F., and Y.D. are supported by the National Natural Science Foundation of China under Grant No. 12275263 and the Quantum Science and Technology-National Science and Technology Major Project under Grant No. 2021ZD0301900. Y.D. is also supported by the Natural Science Foundation of Fujian Province of China under Grant No. 2023J02032. Z.F. is also supported by the NSFC under Grant No. 12504265.
\end{acknowledgments}

\bibliographystyle{apsrev4-2}
\bibliography{references}

\end{document}

\title{Supplemental Material for ``Perturbative Renormalization and Universality Diagram for Long-Range Quantum Criticality''}

\author{Zhiyi Li}
\affiliation{Department of Modern Physics, University of Science and Technology of China, Hefei, Anhui 230026, China}
\affiliation{Hefei National Laboratory, University of Science and Technology of China, Hefei 230088, China}

\author{Zhijie Fan}
\affiliation{Hefei National Laboratory, University of Science and Technology of China, Hefei 230088, China}
\affiliation{Shanghai Research Center for Quantum Science and CAS Center for Excellence in Quantum Information and Quantum Physics, University of Science and Technology of China, Shanghai 201315, China}
\affiliation{Hefei National Research Center for Physical Sciences at the Microscale and School of Physical Sciences, University of Science and Technology of China, Hefei 230026, China}

\author{Kun Chen}
\email{chenkun@itp.ac.cn}
\affiliation{CAS Key Laboratory of Theoretical Physics, Institute of Theoretical Physics, Chinese Academy of Sciences, Beijing 100190, China}

\author{Youjin Deng}
\email{yjdeng@ustc.edu.cn}
\affiliation{Department of Modern Physics, University of Science and Technology of China, Hefei, Anhui 230026, China}
\affiliation{Hefei National Laboratory, University of Science and Technology of China, Hefei 230088, China}
\affiliation{Hefei National Research Center for Physical Sciences at the Microscale and School of Physical Sciences, University of Science and Technology of China, Hefei 230026, China}

\date{\today}

\maketitle

\section{Tree-Level Scaling and Phase Boundaries}

Throughout this Supplemental Material, long-range and short-range are abbreviated as LR and SR, respectively.
We use the continuum action and normalization conventions of the main text.
Here and below, $k$ denotes the magnitude of the momentum vector, $k=|\mathbf{k}|$.
The full bare inverse propagator is
\begin{equation}
G_0^{-1}(\omega,k)=\omega^2+|k|^\sigma+Kk^2+r .
\label{eq:sm_full_prop}
\end{equation}
In the LR regime $\sigma<2$, the analytic $Kk^2$ term is irrelevant and will be dropped unless explicitly stated.
The LR inverse propagator is therefore
\begin{equation}
G_0^{-1}(\omega,k)=\omega^2+|k|^\sigma+r
\label{eq:sm_lr_prop_inv}
\end{equation}
and fixes
\begin{equation}
2z=\sigma,\qquad z_0=\frac{\sigma}{2}.
\label{eq:sm_z0}
\end{equation}
We use
\begin{equation}
\int_{\omega,k}\equiv
\int\frac{d\omega}{2\pi}\frac{d^d k}{(2\pi)^d}.
\label{eq:sm_measure}
\end{equation}
The critical two-point vertex is parameterized as
\begin{equation}
\Gamma_*^{(2)}(0,k)\sim|k|^{2-\eta_k};~~~~~~ \Gamma_*^{(2)}(\omega,0)\sim \omega^{2-\eta_\omega}.
\label{eq:sm_vertex_def}
\end{equation}
Since the bare LR kernel is $|k|^\sigma=|k|^{2-\delta}$ with $\delta=2-\sigma$, the tree-level anomalous dimensions are
\begin{equation}
\eta_k^{(0)}=\delta,\qquad \eta_\omega^{(0)}=0.
\label{eq:sm_tree_eta}
\end{equation}
The dynamical exponent follows by matching the frequency and momentum terms:
\begin{equation}
z=\frac{2-\eta_k}{2-\eta_\omega}.
\label{eq:sm_z_def}
\end{equation}
The quadratic action gives
\begin{equation}
2[\phi]+\sigma-(d+z)=0,\qquad
[\phi]=\frac{d+z-\sigma}{2}
=\frac{d-\sigma/2}{2}.
\label{eq:sm_phi_dim}
\end{equation}
The mass eigenvalue is $y_r^{\rm LR}=\sigma$, so the LR Gaussian correlation-length exponent is $\nu_0^{-1}=\sigma$.
At the LR Gaussian fixed point, the quartic coupling has tree-level eigenvalue
\begin{equation}
y_u^{\rm LR}
=d+z-4[\phi]
=2\sigma-(d+z)
=\frac{3\sigma}{2}-d.
\label{eq:sm_yu}
\end{equation}
Therefore the LR upper critical surface is
\begin{equation}
d_u=\frac{3\sigma}{2},
\qquad
\sigma=\frac{2d}{3}.
\label{eq:sm_upper}
\end{equation}
The analytic SR gradient has relative eigenvalue
\begin{equation}
y_{k^2}^{\rm LR}=\sigma-2.
\label{eq:sm_k2}
\end{equation}
It is irrelevant for $\sigma<2$ and becomes marginal at $\sigma=2$.

The lower critical boundary follows from the infrared fluctuation of the Gaussian critical theory.
At $r=0$,
\begin{equation}
\langle\phi^2\rangle
\sim\int d^d k\,d\omega\,\frac{1}{\omega^2+|k|^\sigma}
\sim\int_0^\Lambda dk\,k^{d-1-\sigma/2}.
\label{eq:sm_lcd_integral}
\end{equation}
This integral is infrared convergent only for $d>\sigma/2$.
Thus the LR lower critical dimension is
\begin{equation}
d_\ell^{\rm LR}=\frac{\sigma}{2},
\qquad
\sigma=2d .
\label{eq:sm_lower}
\end{equation}

In our work, we expand our theory with $d = 3-\epsilon$ and denotes $\sigma=2-\delta$, 
the non-mean-field LR window is thus consider as 
$0<\delta<\frac{2\epsilon}{3}$, as the whole interacting LR region near $d=3$ is accessible within our expansion.

\section{One-Loop Fixed Point}

For $\sigma<2$, the free propagator is regarded as 
\begin{equation}
G_0(\omega,k)=\frac{1}{\omega^2+|k|^\sigma+r}.
\label{eq:sm_prop}
\end{equation}
We repeatedly use the integral
\begin{equation}
\int_{-\infty}^{\infty}\frac{d\omega}{2\pi}
\frac{1}{(\omega^2+E^2)^\alpha}
=\frac{1}{2\sqrt{\pi}}
\frac{\Gamma(\alpha-1/2)}{\Gamma(\alpha)}
E^{1-2\alpha}.
\label{eq:sm_freq_general}
\end{equation}
In particular,
\begin{equation}
\int\frac{d\omega}{2\pi}\frac{1}{\omega^2+E^2}
=\frac{1}{2E},\qquad
\int\frac{d\omega}{2\pi}\frac{1}{(\omega^2+E^2)^2}
=\frac{1}{4E^3}.
\label{eq:sm_freq_special}
\end{equation}

The one-loop four-point bubble is controlled by
\begin{equation}
I_2(r)=
\int_{\omega,k}\frac{1}{(\omega^2+|k|^\sigma+r)^2}
=\frac{1}{4}\int_k (|k|^\sigma+r)^{-3/2}.
\label{eq:sm_bubble_start}
\end{equation}
Using spherical coordinates and $t=|k|^\sigma/r$,
\begin{align}
I_2(r)
&=\frac{S_d}{(2\pi)^d}\frac{1}{4\sigma}
r^{d/\sigma-3/2}
B\!\left(\frac{d}{\sigma},\frac{3}{2}-\frac{d}{\sigma}\right)
\nonumber\\
&=\frac{S_d}{(2\pi)^d}\frac{1}{4\sigma}
r^{d/\sigma-3/2}
\frac{\Gamma(d/\sigma)\Gamma(3/2-d/\sigma)}{\Gamma(3/2)} ,
\label{eq:sm_bubble_beta}
\end{align}
where $S_d=2\pi^{d/2}/\Gamma(d/2)$.
It is useful to denote
\begin{equation}
y\equiv \epsilon-\frac{3}{2}\delta .
\label{eq:sm_ydef_one_loop}
\end{equation}
Therefore, the UV pole part is
\begin{equation}
I_2(r)\big|_{\rm pole}
=
\frac{S_d}{(2\pi)^d}\frac{1}{4}
\frac{\Gamma(d/\sigma)}{\Gamma(3/2)}
\frac{1}{\epsilon-3\delta/2}
+O(1).
\label{eq:sm_bubble_pole}
\end{equation}

We define the dimensionless coupling by absorbing this residue:
\begin{equation}
g=C_{\sigma,d}\,\mu^{-(\epsilon-3\delta/2)}u,
\qquad
C_{\sigma,d}
=\frac{S_d}{(2\pi)^d}\frac{1}{4}
\frac{\Gamma(d/\sigma)}{\Gamma(3/2)} .
\label{eq:sm_coupling}
\end{equation}
With the standard $O(n)$ combinatorial factor, the relation between the bare coupling and the renormalized coupling is
\begin{equation}
u_0
=\mu^y C_{\sigma,d}^{-1}
\left[
g+\frac{n+8}{6}\frac{g^2}{y}
+O(g^3)
\right],
\label{eq:sm_bare_coupling}
\end{equation}
where $u_0$ is the bare coupling.
Since $u_0$ cannot depend on the arbitrary scale $\mu$,
\begin{equation}
0=\mu\frac{d u_0}{d\mu}
=y\left[
g+\frac{n+8}{6}\frac{g^2}{y}
\right]
+
\beta_g\left[
1+\frac{n+8}{3}\frac{g}{y}
\right]
+O(g^3).
\label{eq:sm_beta_from_u0}
\end{equation}
Solving this equation order by order gives
\begin{equation}
\beta_g\equiv \mu\frac{d g}{d\mu}
=-\left(\epsilon-\frac{3}{2}\delta\right)g
+\frac{n+8}{6}g^2+O(g^3).
\label{eq:sm_beta}
\end{equation}
The LR Wilson-Fisher (LR-WF) fixed point is therefore given by solving $\beta_g(g^*) = 0$ with
\begin{equation}
g_*=\frac{6}{n+8}
\left(\epsilon-\frac{3}{2}\delta\right)+O(\epsilon^2).
\label{eq:sm_gstar}
\end{equation}

The mass beta function follows from the tadpole:
\begin{equation}
I_1(r)=
\int_{\omega,k}\frac{1}{\omega^2+|k|^\sigma+r}
=\frac{S_d}{(2\pi)^d}\frac{1}{2\sigma}
r^{d/\sigma-1/2}
\frac{\Gamma(d/\sigma)\Gamma(1/2-d/\sigma)}{\Gamma(1/2)} .
\label{eq:sm_tadpole}
\end{equation}
The pole part of this integral renormalizes the coefficient of $\phi_a^2$, and the $O(n)$ contraction of the tadpole in the $\frac{u}{4!}(\phi_a\phi_a)^2$ theory gives the factor $(n+2)/6$.
Together with the canonical dimension $[r]=\sigma$, this yields
\begin{equation}
\beta_r
=-\left[
\sigma-\frac{n+2}{6}g+O(g^2)
\right]r .
\label{eq:sm_beta_r}
\end{equation}
Therefore
\begin{equation}
\frac{1}{\nu} = y_r= - \frac{d\beta_r}{dr}
=\sigma-\frac{n+2}{6}g_*+O(\epsilon^2)
=2-\delta-\frac{n+2}{n+8}
\left(\epsilon-\frac{3}{2}\delta\right)+O(\epsilon^2).
\label{eq:sm_nu}
\end{equation}
Equation~\eqref{eq:sm_nu} therefore fixes the correlation-length exponent.

\section{Two-Loop Sunset Integral: Frequency Convolution}

Kinetic anomalous dimensions first appear at two loops.
The one-loop tadpole is independent of the external frequency and momentum.
We denote the external frequency by $\Omega$ to distinguish it from the internal frequencies $\omega_1$ and $\omega_2$.
The two-loop sunset self-energy is
\begin{equation}
\Sigma^{(2)}(\Omega,k)
=-\frac{n+2}{18}u^2
\int_{\omega_1,p}\int_{\omega_2,q}
G_1G_2G_3,
\label{eq:sm_sunset}
\end{equation}
where
\begin{equation}
G_1^{-1}=\omega_1^2+E_1^2,\quad
G_2^{-1}=\omega_2^2+E_2^2,\quad
G_3^{-1}=(\Omega+\omega_1+\omega_2)^2+E_3^2,
\end{equation}
and
\begin{equation}
E_1=|p|^{\sigma/2},\qquad
E_2=|q|^{\sigma/2},\qquad
E_3=|p+q+k|^{\sigma/2}.
\label{eq:sm_Edefs}
\end{equation}
Let $E_\Sigma=E_1+E_2+E_3$.
The two internal frequency integrals can be done exactly.
Using
\begin{equation}
f_i(t)\equiv
\int\frac{d\omega}{2\pi}
\frac{e^{i\omega t}}{\omega^2+E_i^2}
=\frac{e^{-E_i|t|}}{2E_i},
\label{eq:sm_fi}
\end{equation}
we write
\begin{align}
I_\omega(\Omega)
&=\int\frac{d\omega_1d\omega_2}{(2\pi)^2}
G_1(\omega_1)G_2(\omega_2)G_3(\Omega+\omega_1+\omega_2)
\nonumber\\
&=\int dt_1dt_2dt_3\,
f_1(t_1)f_2(t_2)f_3(t_3)e^{-i\Omega t_3}
\nonumber\\
&\quad\times
\int\frac{d\omega_1}{2\pi}e^{-i\omega_1(t_1+t_3)}
\int\frac{d\omega_2}{2\pi}e^{-i\omega_2(t_2+t_3)} .
\label{eq:sm_conv_steps}
\end{align}
The internal frequency integrals impose $t_1=t_2=-t_3$.
Since each $f_i(t)$ is even,
\begin{equation}
I_\omega(\Omega)
=\int_{-\infty}^{\infty}dt\,
e^{-i\Omega t}
\frac{e^{-E_\Sigma |t|}}{8E_1E_2E_3}.
\end{equation}
Using
\begin{equation}
\int_{-\infty}^{\infty}dt\,e^{-i\Omega t}e^{-E_\Sigma |t|}
=\frac{2E_\Sigma}{E_\Sigma^2+\Omega^2},
\end{equation}
we obtain
\begin{equation}
\int\frac{d\omega_1d\omega_2}{(2\pi)^2}
\prod_{i=1}^3G_i
=
\frac{E_\Sigma}{4E_1E_2E_3(E_\Sigma^2+\Omega^2)}.
\label{eq:sm_freq_conv}
\end{equation}

\section{Frequency Kinetic Counterterm}

The coefficient of $\Omega^2$ follows from expanding Eq.~\eqref{eq:sm_freq_conv}:
\begin{equation}
\frac{E_\Sigma}{4E_1E_2E_3(E_\Sigma^2+\Omega^2)}
=\frac{1}{4E_1E_2E_3E_\Sigma}
-\frac{\Omega^2}{4E_1E_2E_3E_\Sigma^3}
+O(\Omega^4).
\end{equation}
Thus
\begin{equation}
\frac{\partial\Sigma^{(2)}(\Omega,0)}{\partial\Omega^2}\bigg|_{\Omega=0}
=\frac{n+2}{18}u^2
\int_{p,q}\frac{1}{4E_1E_2E_3E_\Sigma^3}.
\label{eq:sm_freq_deriv}
\end{equation}
For $k=0$, rescale $p=\rho x$, $q=\rho y$.
Then
\begin{equation}
d^dp\,d^dq=\rho^{2d-1}d\rho\,d\Omega_{2d-1},
\qquad
E_i=\rho^{\sigma/2}e_i(x,y),
\end{equation}
and the denominator in Eq.~\eqref{eq:sm_freq_deriv} has degree $3\sigma$.
Therefore the UV radial integral scales as
\begin{equation}
\int^\infty d\rho\,\rho^{2d-1-3\sigma}.
\label{eq:sm_freq_radial}
\end{equation}
In dimensional regularization this logarithmic divergence is represented by
\begin{equation}
\Gamma\!\left(\frac{3\sigma-2d}{2}\right)
=\Gamma\!\left(\frac{2\epsilon-3\delta}{2}\right)
=\frac{2}{2\epsilon-3\delta}+O(1).
\label{eq:sm_freq_pole}
\end{equation}
This is the pole that fixes the frequency kinetic counterterm.

\section{Momentum Fractional-Kernel Counterterm}

The momentum kinetic counterterm is extracted from the coefficient of the nonanalytic operator $|k|^\sigma$.
It comes from the first term in the small-$\Omega$ expansion,
\begin{equation}
F(k;p,q)=\frac{1}{4E_1E_2E_3E_\Sigma},
\qquad E_3=|p+q+k|^{\sigma/2}.
\label{eq:sm_F}
\end{equation}
Let
\begin{equation}
X\equiv |k|^\sigma,\qquad
k=X^{1/\sigma}\hat k ,
\label{eq:sm_X}
\end{equation}
with fixed direction $\hat k$.
For $P=p+q$,
\begin{equation}
\frac{\partial E_3}{\partial X}
=\frac{\partial}{\partial X}|P+X^{1/\sigma}\hat k|^{\sigma/2}
=\frac{1}{2}X^{1/\sigma-1}|P+k|^{\sigma/2-2}(P+k)\cdot\hat k .
\label{eq:sm_dE}
\end{equation}
Differentiating $F$ gives
\begin{equation}
\frac{\partial F}{\partial X}
=-\frac{1}{4E_1E_2}
\frac{\partial E_3}{\partial X}
\left(
\frac{1}{E_3^2E_\Sigma}
+\frac{1}{E_3E_\Sigma^2}
\right).
\label{eq:sm_dF}
\end{equation}
The two terms in parentheses come respectively from differentiating $E_3^{-1}$ and $E_\Sigma^{-1}$.
For a common large internal momentum scale $Q$, the energies scale as $E_1,E_2,E_3,E_\Sigma\sim Q^{\sigma/2}$.
The leading part of Eq.~\eqref{eq:sm_dE} scales as $Q^{\sigma/2-1}\hat P\cdot\hat k$.
Combined with the factors in Eq.~\eqref{eq:sm_dF}, it gives a term of order $Q^{-2\sigma-1}\hat P\cdot\hat k$.
This term is odd under angular averaging and cannot contribute to the rotationally invariant two-point vertex.
The first nonzero scalar contribution is obtained by expanding the large-$Q$ integrand one order further in the external direction, which supplies one additional power of $1/Q$.
Consequently, after projection onto the $|k|^\sigma$ operator, the integrand has homogeneous degree $-2\sigma-2$.
The radial integral therefore scales as
\begin{equation}
\int^\infty d\rho\,\rho^{2d-1-2\sigma-2}.
\label{eq:sm_k_radial}
\end{equation}
The corresponding dimensional pole is
\begin{equation}
\Gamma\!\left(\frac{2\sigma+2-2d}{2}\right)
=\Gamma(\epsilon-\delta)
=\frac{1}{\epsilon-\delta}+O(1)
=\frac{2}{2\epsilon-2\delta}+O(1).
\label{eq:sm_k_pole}
\end{equation}
Thus the frequency and momentum kinetic counterterms have different pole denominators:
\begin{equation}
2\epsilon-3\delta
\quad\text{and}\quad
2\epsilon-2\delta .
\label{eq:sm_two_poles}
\end{equation}

\section{Counterterms and Anomalous Dimensions}

After the mass counterterm has removed the momentum-independent part of the sunset graph, the remaining UV poles are proportional to $\Omega^2$ and $|k|^\sigma$.
We write
\begin{equation}
Z_\omega=1+\delta Z_\omega,\qquad
Z_k=1+\delta Z_k.
\label{eq:sm_Zdefs}
\end{equation}
The renormalized inverse propagator is organized as
\begin{equation}
\Gamma_R^{(2)}(\Omega,k)
=Z_\omega\Omega^2+Z_k|k|^\sigma+r_R+\Sigma_{\rm finite}(\Omega,k)+\cdots .
\label{eq:sm_GammaR}
\end{equation}
Minimal subtraction fixes $\delta Z_\omega$ and $\delta Z_k$ by requiring the coefficients of $\Omega^2$ and $|k|^\sigma$ in $\Gamma_R^{(2)}$ to be finite.

It remains to state how the pole residues are normalized.
Before rewriting the result in terms of $g$, the two kinetic counterterms have the form
\begin{equation}
\delta Z_\omega
=\frac{n+2}{18}u^2 R_\omega^{(u)}
\frac{2}{2\epsilon-3\delta}
+O(u^3),
\qquad
\delta Z_k
=\frac{n+2}{18}u^2 R_k^{(u)}
\frac{2}{2\epsilon-2\delta}
+O(u^3),
\label{eq:sm_unren_residues}
\end{equation}
where $R_\omega^{(u)}$ and $R_k^{(u)}$ denote the finite angular residues after the Gamma-function poles have been isolated.
At the order kept here these residues may be evaluated at $(d,\sigma)=(3,2)$.
For the frequency channel this is the standard massless sunset integral in $D=4-\bar y$ dimensions,
\begin{equation}
J(P)=\int_Q\int_K
\frac{1}{Q^2K^2(Q+K+P)^2},
\qquad \bar y=4-D,
\label{eq:sm_standard_sunset}
\end{equation}
with $\int_Q\equiv\int d^DQ/(2\pi)^D$, and its $P^2$ pole is
\begin{equation}
J(P)\big|_{\rm pole}
=-\frac{P^2}{2(4\pi)^4}\frac{1}{\bar y}+O(\bar y^0).
\label{eq:sm_standard_sunset_pole}
\end{equation}
The sign is absorbed by the self-energy and counterterm convention, so the positive residue entering Eq.~\eqref{eq:sm_unren_residues} is
\begin{equation}
R_\omega^{(u)}
=\frac{1}{2(4\pi)^4}
=\frac{1}{8}C_{3,2}^2,
\qquad
C_{3,2}=\frac{1}{8\pi^2}.
\label{eq:sm_residue_match}
\end{equation}
The nonlocal momentum projection has the same leading normalized residue, $R_k^{(u)}=C_{3,2}^2/8+O(\epsilon,\delta)$; the difference between the two kinetic projections is therefore in the pole denominator, not in the leading residue.
Using $u=\mu^y C_{\sigma,d}^{-1}g$, the factor $\mu^{2y}$ changes only finite terms in minimal subtraction, and Eq.~\eqref{eq:sm_unren_residues} becomes
\begin{equation}
\delta Z_\omega
=\frac{n+2}{72}g^2
\frac{1}{2\epsilon-3\delta}
+O(g^3),
\label{eq:sm_dZw}
\end{equation}
\begin{equation}
\delta Z_k
=\frac{n+2}{72}g^2
\frac{1}{2\epsilon-2\delta}
+O(g^3).
\label{eq:sm_dZk}
\end{equation}
More generally, the residues multiplying these poles are analytic functions of $\epsilon$ and $\delta$ and equal to one at $(\epsilon,\delta)=(0,0)$ in this normalization.
Their $O(\epsilon,\delta)$ corrections contribute only at $O(\epsilon^3)$ to the anomalous dimensions retained here.

The Callan-Symanzik anomalous dimensions are
\begin{equation}
\gamma_\alpha(g)
=-\frac{1}{2}\beta_g\frac{\partial \delta Z_\alpha}{\partial g},
\qquad \alpha=\omega,k.
\label{eq:sm_gamma_def}
\end{equation}
The physical exponents are
\begin{equation}
\eta_\omega=2\gamma_\omega(g_*),\qquad
\eta_k=\delta+2\gamma_k(g_*).
\label{eq:sm_eta_gamma}
\end{equation}
The extra $\delta$ in $\eta_k$ is the canonical contribution of the bare LR kernel.

Let again $y=\epsilon-3\delta/2$.
From Eqs.~\eqref{eq:sm_dZw} and \eqref{eq:sm_dZk},
\begin{equation}
\frac{\partial\delta Z_\omega}{\partial g}
=\frac{n+2}{36}\frac{g}{2\epsilon-3\delta}
=\frac{n+2}{72}\frac{g}{y},
\qquad
\frac{\partial\delta Z_k}{\partial g}
=\frac{n+2}{36}\frac{g}{2\epsilon-2\delta}
=\frac{n+2}{72}\frac{g}{\epsilon-\delta}.
\label{eq:sm_dZ_derivatives}
\end{equation}
In the present order only the canonical part $\beta_g=-yg+O(g^2)$ is needed in Eq.~\eqref{eq:sm_gamma_def}, because the counterterms already start at $g^2$.
Therefore
\begin{equation}
2\gamma_\omega(g)
=\frac{n+2}{72}g^2+O(g^3),
\qquad
2\gamma_k(g)
=\frac{n+2}{72}g^2\frac{y}{\epsilon-\delta}+O(g^3).
\label{eq:sm_gamma_intermediate}
\end{equation}
Evaluating these expressions at Eq.~\eqref{eq:sm_gstar} gives
\begin{equation}
\eta_\omega
=\frac{n+2}{2(n+8)^2}
\left(\epsilon-\frac{3}{2}\delta\right)^2
+O(\epsilon^3),
\label{eq:sm_etaomega}
\end{equation}
\begin{equation}
\eta_k
=\delta+
\frac{n+2}{2(n+8)^2}
\frac{\left(\epsilon-\frac{3}{2}\delta\right)^3}{\epsilon-\delta}
+O(\epsilon^3).
\label{eq:sm_etak}
\end{equation}

\section{Critical Exponents and Limiting Checks}

Combining the one-loop mass scaling and the two-loop kinetic renormalization gives
\begin{equation}
\frac{1}{\nu}
=2-\delta
-\frac{n+2}{n+8}
\left(\epsilon-\frac{3}{2}\delta\right)
+O(\epsilon^2),
\label{eq:sm_final_nu}
\end{equation}
\begin{equation}
\eta_\omega
=\frac{n+2}{2(n+8)^2}
\left(\epsilon-\frac{3}{2}\delta\right)^2
+O(\epsilon^3),
\label{eq:sm_final_etaomega}
\end{equation}
\begin{equation}
\eta_k
=\delta+
\frac{n+2}{2(n+8)^2}
\frac{\left(\epsilon-\frac{3}{2}\delta\right)^3}{\epsilon-\delta}
+O(\epsilon^3).
\label{eq:sm_final_etak}
\end{equation}
The dynamical exponent is
\begin{equation}
z=\frac{2-\eta_k}{2-\eta_\omega}.
\label{eq:sm_final_zdef}
\end{equation}
To the same order,
\begin{align}
z
&=1-\frac{\delta}{2}
+\frac{1}{2}\left[\eta_\omega-(\eta_k-\delta)\right]
+O(\epsilon^3)
\nonumber\\
&=1-\frac{\delta}{2}
+\frac{n+2}{4(n+8)^2}
\left[
\left(\epsilon-\frac{3}{2}\delta\right)^2
-\frac{\left(\epsilon-\frac{3}{2}\delta\right)^3}{\epsilon-\delta}
\right]
+O(\epsilon^3)
\nonumber\\
&=1-\frac{\delta}{2}
+\frac{(n+2)\delta}{8(n+8)^2(\epsilon-\delta)}
\left(\epsilon-\frac{3}{2}\delta\right)^2
+O(\epsilon^3).
\label{eq:sm_final_z}
\end{align}

At the SR boundary, $\delta=0$ and $\sigma=2$.
Equations~\eqref{eq:sm_final_etaomega} and \eqref{eq:sm_final_etak} give
\begin{equation}
\eta_\omega=\eta_k
=\frac{n+2}{2(n+8)^2}\epsilon^2+O(\epsilon^3),
\qquad z=1.
\label{eq:sm_sr_check}
\end{equation}
This is the relativistic SR quantum Wilson-Fisher limit.
At the LR mean-field boundary, $\delta=2\epsilon/3$, so $g_*=0$ and all loop corrections vanish:
\begin{equation}
\eta_\omega=0,\qquad
\eta_k=\delta=2-\sigma,\qquad
z=\frac{\sigma}{2}.
\label{eq:sm_mf_check}
\end{equation}
This reproduces the LR Gaussian boundary.